\documentclass[12pt]{article}
\pdfoutput=1
\usepackage{graphicx}
\usepackage{epstopdf}
\usepackage{amsmath}
\usepackage{amsfonts}
\usepackage{amssymb}
\usepackage{color}
\usepackage{mathrsfs}
\usepackage{float}

\newcommand{\be}{\begin{equation}}
\newcommand{\ee}{\end{equation}}
\newcommand{\barr}{\begin{array}}
\newcommand{\earr}{\end{array}}

\usepackage{latexsym}
\usepackage{amsthm}
\usepackage{amsmath}
\usepackage{graphicx}
\usepackage{amssymb}
\usepackage{bbm}
\newcommand{\gsim}{\lower.7ex\hbox{$\;\stackrel{\textstyle>}{\sim}\;$}}
\newcommand{\lsim}{\lower.7ex\hbox{$\;\stackrel{\textstyle<}{\sim}\;$}}

\newcommand{\bea}{\begin{eqnarray}}
\newcommand{\eea}{\end{eqnarray}}

\newcommand{\comment}[1]{}

\def\e{\mathrm{e}}

\def\half{{1\over 2}}

\def\half{{1\over 2}}
\def\({\left(}
\def\){\right)}
\def\[{\left[}
\def\]{\right]}
\def\e{\begin{equation}}
\def\q{\end{equation}}
\def\m{\begin{eqnarray}}
\def\n{\end{eqnarray}}

\begin{document}


\setcounter{page}{1} \baselineskip=15.5pt \thispagestyle{empty}

\begin{flushright}
\end{flushright}
\vfil

\begin{center}

{\Large \bf Distance Priors from \emph{Planck} 2015 data}
\\[0.7cm]
{Qing-Guo Huang$^\heartsuit$ \footnote{huangqg@itp.ac.cn}, Ke Wang$^{\heartsuit,\ \Diamond}$ \footnote{wangke@itp.ac.cn} and Sai Wang$^{\heartsuit}$ \footnote{wangsai@itp.ac.cn}}
\\[0.7cm]

{\normalsize { \sl $^\heartsuit$ State Key Laboratory of Theoretical Physics, Institute of Theoretical Physics, \\ Chinese Academy of Science, Beijing 100190, China}}\\
\vspace{.2cm}

{\normalsize { \sl $^\Diamond$  University of the Chinese Academy of Sciences, Beijing 100190, China}}
\vspace{.3cm}

\end{center}

\vspace{.8cm}

\hrule \vspace{0.3cm}
{\small  \noindent \textbf{Abstract} \\[0.3cm]
\noindent
}
We update the distance priors by adopting $Planck~ \textrm{TT,TE,EE}+\textrm{lowP}$ data released in 2015, and our results impose at least $30\%$ tighter constraints than those from $Planck~ \textrm{TT}+\textrm{lowP}$. Combining the distance priors with the combination of supernova Union~2.1 compilation of 580 SNe (Union~2.1) and low redshift Baryon Acoustic Oscillation (BAO) data, we constrain the cosmological parameters in the freely binned dark energy (FBDE) and FBDE$+\Omega_k$ models respectively, and find that the equations of state of dark energy in both models are consistent with $w=-1$. Furthermore, we show that the tension with the BAO data at $z=2.34$ from Ly$\alpha$ forest (Ly$\alpha$F) auto-correlation and Combined Ly$\alpha$F cannot be relaxed in the FBDE and FBDE$+\Omega_k$ models.

\vspace{0.3cm}
\hrule
\vspace{6cm}

\begin{flushleft}
\end{flushleft}

\vspace{10cm}

\newpage
\section{Introduction}

The current cosmic acceleration \cite{Riess:1998cb,Perlmutter:1998np} of the universe implies that there is a mysterious component inventory in the universe, namely the dark energy (DE). Superbly accurate data are necessary for understanding the nature of DE. Although the usual luminosity-distance method based on supernovae of type Ia (SNIa) can impose certain constraints on DE, there are limitations \cite{Perlmutter:1996ds,Choudhury:2003tj} for SNIa when we consider the nonzero spatial curvature or an evolving dark energy model. Because the measurements of SNIa span a relatively narrow range of redshifts, some other cosmological observations are called for. To impose constraints on DE models that are beyond SNIa's domain, the acoustic peak method is usually taken as a geometric complement. The acoustic peaks can be imprinted onto not only the late-time power spectrum of cosmic microwave background (CMB) anisotropy, but also that of the non-relativistic matter from which the baryon acoustic oscillation (BAO) can be measured. Although the acoustic features are thought as one of the most powerful ways to probe DE, there are also some limitations for them. One limitation is that the effects on the matter correlations are weak. The other is that the evolution of linear density perturbations, including perturbations of both traditional components and DE, should be took into account when we fit a certain DE model to CMB data through the full Boltzmann analysis \cite{Weller:2003hw,Bean:2003fb,Li:2008cj}. However, even though this prerequisite is ready, this method will be prohibitive because of the long computing time.

Compounding the situation is that for several more imaginative DE models, we can't set up the equations for linear density perturbations or we can set up these equations only in principle. DGP model \cite{Dvali:2000hr}, for example, is given first by an action. If one wants to get the CMB temperature power spectrum from scratch using it, one will get stuck with complications of setting up these equations \cite{Koyama:2006ef,Koyama:2005kd,Sawicki:2006jj,Song:2006jk} and treating them numerically. Another hard case is that some DE models are based on phenomenological considerations, and then there is no enough information to calculate the density perturbations. Caedassian model \cite{Gondolo:2002fh,Freese:2002sq}, for example, with a modified Friedmann equation gets its energy density perturbations through the fluid flow approach of Hawking \cite{Lyth:1990bs,9.2}. However, the results are very different when we choose a different gauge, which creates a problem of understanding perturbations in the current Universe.

Fortunately, some pioneers have developed a substitution for full Boltzmann analysis of CMB anisotropy, which can not only incorporate as much empirical information as possible but also avoid the calculation of the evolution of of linear density perturbations. The fundamental principle is that using certain characteristic distance scales to summarize the CMB data, namely the shift parameter $R$ \cite{Bond:1997wr,Efstathiou:1998xx} that determines the amplitude of acoustic peaks in the power spectrum of CMB temperature anisotropy and the acoustic scale $l_A$ that determines the acoustic peak structure \cite{Wang:2007mza}. These two distance scales from CMB are so-called distance priors.
From WMAP \cite{Komatsu:2008hk, Komatsu:2010fb} to \emph{Planck} satellite \cite{Ade:2015rim}, almost every data release was accompanied by a corresponding set of distance priors.

In this paper, we will update the distance priors with the latest CMB data \cite{data}, in particular including high $\ell$ polarizations data, released by \emph{Planck} collaborations recently, and make comparison to the other two sets \cite{Ade:2015rim,Wang:2013mha} derived from \emph{Planck} 2013 data and \emph{Planck} 2015 data excluding high $\ell$ polarizations data.
The rest of this paper is arranged as follows. We describe our methodology and present our results in Sec.~2. We use the distance priors derived in this paper to constrain the equation of state of dark energy in Sec.~3. The conclusion and discussion are given in Sec.~4.

\section{Reconstruct the distance priors from \emph{Planck} 2015 data}

\subsection{Methodology}

Distance priors can be directly derived from public data, such as \emph{Planck} or WMAP data which are gained through a full Boltzmann analysis of CMB data. In order to obtain the distance priors, one should assume a cosmological model first \cite{Elgaroy:2007bv}. It is necessary to show that the distance priors are effective observables. We will see that it is indeed the case later. Moreover, since we usually use distance priors to deal with the late-time expansion history of the universe, we also add the baryon density today $\Omega_b h^2$ to the distance priors, which is useful to probe the late-time universe but not sensitive to the cosmological models.

We derive the distance priors $l_A$ and $R$ by following \cite{Komatsu:2008hk, dark energy book}. The comoving scale of the first acoustic peak is well determined as $\lambda_p =r_s(z_*)/\pi$, where $r_s$ is the comoving sound horizon which is given by
\begin{align}\label{}
r_s(z)  &= \frac{c}{H_0} \int_0^{1/(1+z)} \frac {da} {a^2E(a)\sqrt{3(1+\frac{3\Omega_bh^2}{4\Omega_\gamma h^2}a)}}\ ,\nonumber\\
\frac{3}{4\Omega_\gamma h^2}  &= 31500(T_{\textrm{CMB}}/2.7K)^{-4},~~T_{\textrm{CMB}}= 2.7255K\ .
\end{align}
Here $z_*$ is the redshift to the photo-decoupling surface, which is given by the \textsf{CAMB} package.
If we observe an angle $\theta_{\textrm{A}}$ subtending the transverse comoving scale $\lambda_p$, the angular diameter distance is given as $D_\textrm{A}(z_*)=\frac{\lambda_p}{(1+z_*)\theta_{\textrm{A}}}$. It's obvious that different cosmological models will give different $\lambda_p$ and $D_\textrm{A}(z_*)$, but $(1+z_*)$ and $\theta_{\textrm{A}}$ are observables. Therefore, we can easily construct an important derived parameter, i.e. the acoustic scale $l_\textrm{A}$ given by
\begin{align}\label{la}
l_\textrm{A}\equiv\frac{1}{\theta_\textrm{A}}=(1+z_*)\frac{\pi D_\textrm{A}(z_*)}{r_s(z_*)}\ ,
\end{align}
which is an effective observable. Here we used the angular diameter distance
\begin{align}\label{}
D_\textrm{A}
=\frac{c}{1+z}H_0^{-1}  |\Omega_{k}|^{-1/2}\textrm{sinn}\left[|\Omega_{k}|^{1/2}\int_0^z \frac {dz'} {E(z')}\right]\ ,
\end{align}
where $\textrm{sinn}(x)=\textrm{sin}(x)$, $x$, $\textrm{sinh}(x)$ for $\Omega_{k}<0$, $\Omega_{k}=0$, $\Omega_{k}>0$. Here $E(z)$ is given by $E(z)={H(z)}/{H_0}$, i.e.
\begin{align}\label{}
E(z) &= \[ \Omega_{r}(1+z)^4+\Omega_m(1+z)^3+\Omega_{k}(1+z)^2+\Omega_{de} \frac {\rho_{de}(z)} {\rho_{de}(0)} \]^\half\ ,
\end{align}
where $\Omega_{r}$ is the present fractional radiation density
\begin{align}\label{}
\Omega_{r}=\frac{\Omega_{m}}{1+z_{\textrm{eq}}}, ~~z_{\textrm{eq}}=2.5\times10^{4}\Omega_{m}h^2\left(T_{CMB}/2.7\textrm{K}\right)^{-4}\ .
\end{align}
For $\Lambda$CDM and $w$CDM models, $ {\rho_{de}(z)} /{\rho_{de}(0)}$ equals $1$ and $(1+z)^{3(1+w)}$, respectively.

Similarly, we can work out the other important derived parameter, i.e. `shift parameter' $R$, which is also an effective observable. At low redshift, a comoving scale extending along the line of sight spanning $\textrm{d}z$ is given by $\lambda=c\textrm{d}z/H(z)$. It's obvious that different cosmological models will give different $\lambda$ and $H(z)$, but $\textrm{d}z$ is an observable. Therefore, we get a variable $\textrm{d}z=\frac{\lambda H(z)}{c}$.
If we apply it to the decoupling epoch, we can construct
\begin{align}\label{}
\tilde{R}(z_*)=\frac{(1+z_*)D_\textrm{A}(z_*) H(z_*)}{c}\ .
\end{align}
Actually, what we usually use is its another famous version, namely the `shift parameter' $R$ \cite{Bond:1997wr,Efstathiou:1998xx}
\begin{align}\label{Rz}
R(z_*)\equiv\frac{(1+z_*)D_\textrm{A}(z_*) \sqrt{\Omega_m H_0^2}}{c}
\end{align}
That is to say, $R$ is just the traditional construction, but not the only one.

In summary, $l_\textrm{A}$ characterizes the CMB temperature power spectrum in transverse direction, and different $l_\textrm{A}$ gives different distribution of peaks and troughs in the spectrum; $R$ characterizes the CMB temperature power spectrum in line-of-sight direction, and different $R$ will magnify or reduce the amplitude of the acoustic peaks.

\subsection{Results}
Although the distance priors have been derived from \emph{Planck} $\textrm{TT}+\textrm{lowP}$ data in \cite{Ade:2015rim}, the constraints on the distance priors are weaker than those given by Mukherjee et al. in \cite{Mukherjee:2008kd} due to the lack of high $\ell$ polarizations data. Here we derive the distance priors by using the MCMC chains \footnote{From the PLA, we can find four sets of chains named as \\
base\_plikHM\_TTTEEE\_lowTEB, base\_w\_plikHM\_TTTEEE\_lowTEB, \\
base\_omegak\_plikHM\_TTTEEE\_lowTEB and base\_Alens\_plikHM\_TTTEEE\_lowTEB, \\ respectively. These four sets of chains are used in our paper to generate Table 1.} from $Planck~ \textrm{TT,TE,EE}+\textrm{lowP}$ data in \emph{Planck} Legacy Archive (PLA) \cite{data}. The chains of $l_A$ and $R$ can be derived from public MCMC chains by using Eqs.~(\ref{la}) and (\ref{Rz}). Marginalizing over the remaining parameters, we get the mean values and errors of $\{R,l_A,\Omega_bh^2,n_s\}$ as well as their covariance matrix.

\begin{figure}[H]
\begin{center}
\includegraphics[width=10cm]{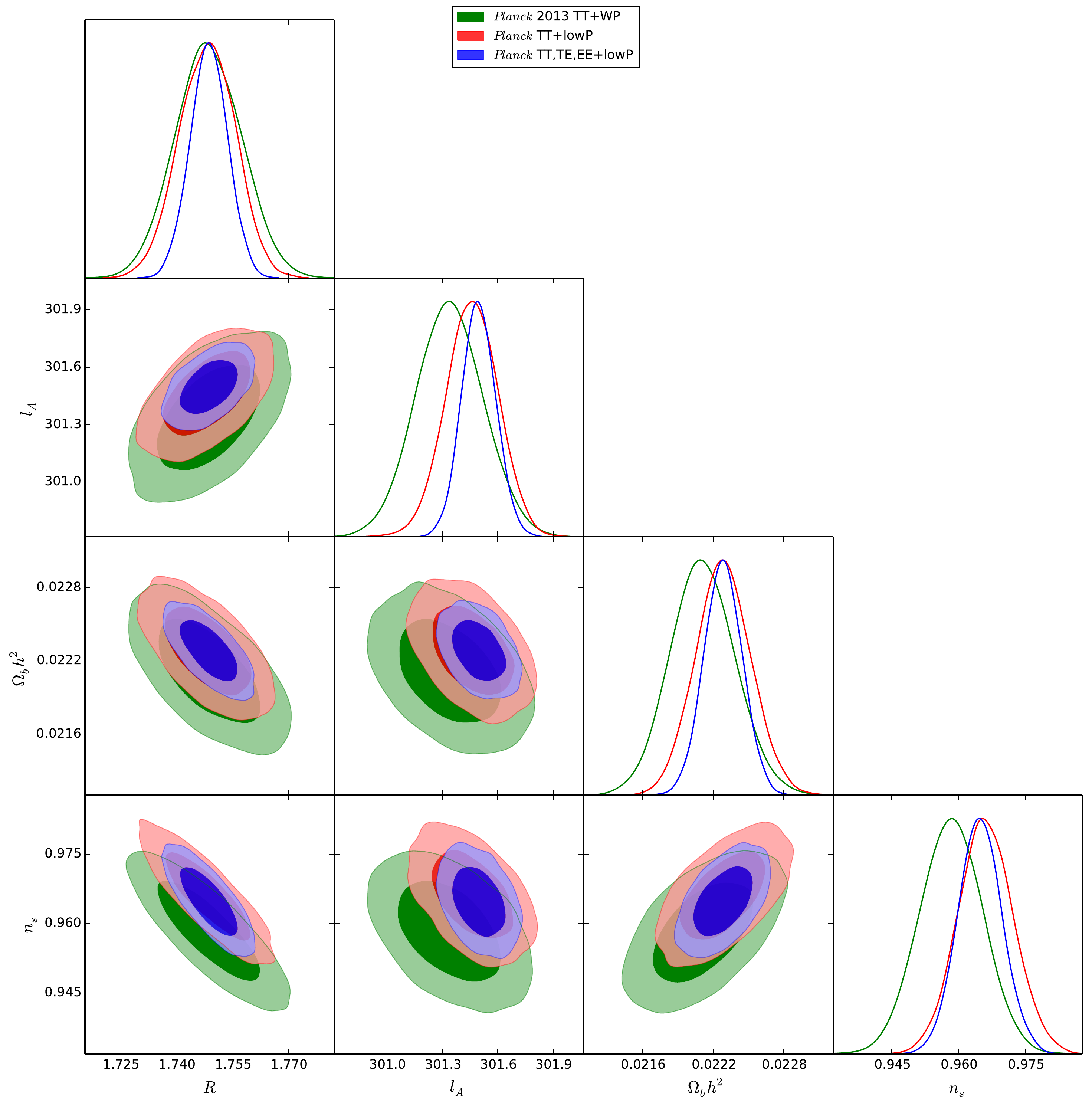}
\end{center}
\caption{Marginalized two-dimension probability contours and one-dimension probability distribution functions of distance priors derived with the $w$CDM model from $Planck$ 2013 \textrm{TT}+\textrm{WP} (green), $Planck~ \textrm{TT}+\textrm{lowP}$ (red) and $Planck~ \textrm{TT,TE,EE}+\textrm{lowP}$ (blue).}
\label{answer_wCDM_tri}
\end{figure}

In Fig.~\ref{answer_wCDM_tri}, there is a comparison among results derived with the $w$CDM model but from different datasets. We find that these three results are consistent with each other, and our results (blue ones) impose at least $30\%$ tighter constraints on the distance priors than those given by \emph{Planck} $\textrm{TT+lowP}$ data.

Meanwhile, in order to explain that the distance priors are effective observables \cite{Mukherjee:2008kd}, we use two different models, i.e. $\Lambda \textrm{CDM}$ and $w$CDM to derive them respectively. The Gaussian likelihood \cite{Wang:2013mha} in $\{R,l_\textrm{A},\Omega_b h^2, n_s\}$ and the covariance matrix are given by the upper two sets in Tab.~\ref{distance priors}. We see that the two sets are stable.


However, due to the smoothing effect of CMB lensing on the power spectrum \cite{Ade:2015xua}, the third set derived in the curved  model (with free $\Omega_k$) is different from the upper two sets by 1$\sigma$. Thus, we provide another set of distance priors which are derived with free $A_L$ (marginalized over later). Here $A_L$ denotes the amplitude of the lensing power spectrum. This set of distance priors is more conservative and used throughout our following discussions.

\begin{table*}[!htp]
\centering
\renewcommand{\arraystretch}{1.5}
\begin{tabular}{cccccc}
\hline\hline
  $\Lambda$CDM & $Planck~ \textrm{TT,TE,EE}+\textrm{lowP}$ & $R$ & $l_\textrm{A}$ & $\Omega_b h^2$ &$n_s$ \\
\hline
$R$                                           & $1.7496\pm0.0050$       & $1.0$&    $0.49$&   $-0.69$&   $-0.77$ \\
$l_\textrm{A}$                                & $301.505\pm0.092$        & $0.49$&    $1.0$&   $-0.37$&   $-0.38$ \\
$\Omega_b h^2$                              & $0.02225\pm0.00016$      & $-0.69$&   $-0.37$&  $1.0$&    $0.51$\\
$n_s$                                        & $0.9646\pm0.0049$       & $-0.77$&   $-0.38$&  $0.51$&    $1.0$\\
\hline\hline
  $w$CDM & $Planck~ \textrm{TT,TE,EE}+\textrm{lowP}$ & $R$ & $l_\textrm{A}$ & $\Omega_b h^2$ &$n_s$ \\
\hline
$R$                                           & $1.7488\pm0.0049$       & $1.0$&    $0.49$&   $-0.68$&   $-0.78$ \\
$l_\textrm{A}$                                & $301.498\pm0.091$         & $0.49$&    $1.0$&   $-0.38$&   $-0.37$ \\
$\Omega_b h^2$                               & $0.02228\pm0.00016$      & $-0.68$&   $-0.38$&  $1.0$&    $0.52$ \\
$n_s$                                        & $0.9648\pm0.0048$       & $-0.78$&   $-0.37$&  $0.52$&    $1.0$\\
\hline\hline
   $\Lambda$CDM$+\Omega_k$ & $Planck~ \textrm{TT,TE,EE}+\textrm{lowP}$ & $R$ & $l_\textrm{A}$ & $\Omega_b h^2$ &$n_s$\\
\hline
$R$                                           & $1.7449\pm0.0052$       & $1.0$&    $0.47$&   $-0.71$&   $-0.79$ \\
$l_\textrm{A}$                                & $301.465\pm0.093$        & $0.47$&    $1.0$&   $-0.37$&   $-0.36$\\
$\Omega_b h^2$                              & $0.02241\pm0.00017$      & $-0.71$&   $-0.37$&  $1.0$&    $0.54$ \\
$n_s$                                        & $0.9679\pm0.0048$       & $-0.79$&   $-0.36$&  $0.54$&    $1.0$\\
\hline\hline
   $\Lambda$CDM$+A_\textrm{L}$ & $Planck~ \textrm{TT,TE,EE}+\textrm{lowP}$ & $R$ & $l_\textrm{A}$ & $\Omega_b h^2$ &$n_s$\\
\hline
$R$                                           & $1.7448\pm0.0054$       & $1.0$&    $0.53$&   $-0.73$&   $-0.80$\\
$l_\textrm{A}$                                & $301.460\pm0.094$        & $0.53$&    $1.0$&   $-0.42$&   $-0.43$ \\
$\Omega_b h^2$                              & $0.02240\pm0.00017$      & $-0.73$&   $-0.42$&  $1.0$&    $0.59$ \\
$n_s$                                        & $0.9680\pm0.0051$       & $-0.80$&   $-0.43$&  $0.59$&    $1.0$ \\
\hline
\end{tabular}
\caption{Distance priors in different cosmological models from $Planck~ \textrm{TT,TE,EE}+\textrm{lowP}$. Here we also list the scalar spectral index $n_s$, it will be useful if one wants to deal with the matter power spectrum.}
\label{distance priors}
\end{table*}

\begin{figure}[H]
\begin{center}
\includegraphics[width=10cm]{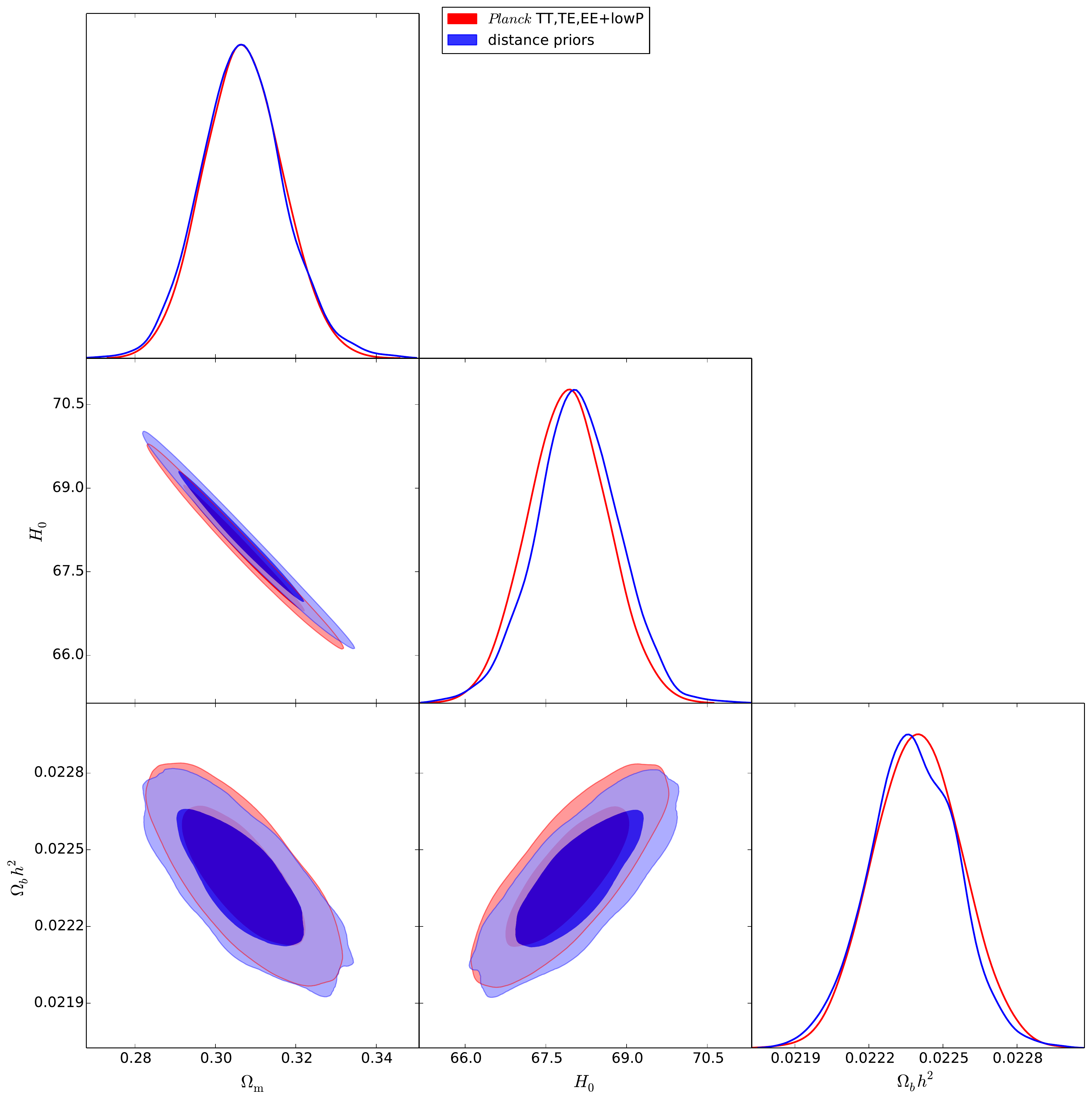}
\end{center}
\caption{Marginalized constraints on parameters in the $\Lambda$CDM model. Contours given by distance priors only are the blue ones, and contours given by $Planck~ \textrm{TT,TE,EE}+\textrm{lowP}$ are the red ones.}
\label{base_Alens_plikHM_TTTEEE_lowTEB_tri}
\end{figure}

To check our distance priors, we constrain the cosmological parameters in the $\Lambda$CDM model by using the distance priors,  and compare them with those constrained by the global fitting from the \emph{Planck} $\textrm{TT,TE,EE+lowP}$ data. The $\chi^2_{\rm distance\ priors}$ is given by
\begin{align}\label{}
\chi^2_{\rm distance\ priors}=\sum(x_i-d_i)(C^{-1})_{ij}(x_j-d_j),
\end{align}
where $x_i=\{R(z_*), l_\textrm{A}(z_*), \Omega_b h^2\}$ are values predicted by $\Lambda$CDM, $d_i=\{R^{Planck}, l_\textrm{A}^{Planck}, {\Omega_b h^2} ^{Planck}\}$, and $C_{ij}$ ($(C^{-1})_{ij}$ is the inverse) is given by the fourth set in Tab.~\ref{distance priors}.
Here we use an approximate expression of $z_*$ to calculate $x_i$, \cite{Hu:1995en},
\e
z_* = 1048[1+0.00124(\Omega_b h^2)^{-0.738}][1+g_1(\Omega_m h^2)^{g_2}]\ ,
\q
where
\m
g_1 &=& \frac{0.0738(\Omega_b h^2)^{-0.238}}{1+39.5(\Omega_b h^2)^{0.763}}\ ,\\
g_2 &=& \frac{0.560}{1+21.1(\Omega_b h^2)^{1.81}}\ .
\n
The two constraints on the parameters $\{\Omega_m, H_0, \Omega_b h^2\}$ from the distance priors and the global fitting are showed in Fig.~\ref{base_Alens_plikHM_TTTEEE_lowTEB_tri}. We find that there are almost complete overlaps between these two sets. Our results indicate that the distance priors can be taken as a good substitute for the global fitting (marginalized over $A_\textrm{L}$) for $\Lambda$CDM model.

\section{Analysis}

\subsection{Dark Energy Model}
Although distance priors can be used to constrain the cosmological parameters in $\Lambda$CDM model, they can't constrain the evolving DE model well, such as $w$CDM and CPL, because DE only plays a crucial role at low redshifts. In order to get a better constraint on the equation of state (EOS) of DE, we need to combine some other low redshift observations, such as the combination of supernova Union~2.1 compilation of 580 SNe (Union~2.1) \cite{Suzuki:2011hu} and low redshift Baryon Acoustic Oscillation (BAO) which are listed in Tab.~\ref{bao}.
\begin{table*}[!htp]
\centering
\renewcommand{\arraystretch}{1.5}
\begin{tabular}{cccc}
\hline\hline
   $z_{\textrm{eff}}$ & measurement & name & reference \\
\hline
$0.106$ &$r_s(z_d)/D_V=0.336\pm0.015$ & 6DFGS &\cite{Beutler:2011hx} \\
$0.15$ & $D_V/r_s(z_d)=(664\pm25)/152.66$ & MGS &\cite{Ross:2014qpa} \\
$0.32$ & $D_V/r_s(z_d)=(1264\pm25)/153.19$ &BOSS LOWZ & \cite{Anderson:2013zyy}\\
$0.57$ & $D_V/r_s(z_d)=(2056\pm20)/153.19$ &BOSS CMASS&\cite{Anderson:2013zyy}\\
\hline
\end{tabular}
\caption{The low redshift BAO data. Here the $z_d$ is given by Eisenstein \& Hu formula.}
\label{bao}
\end{table*}
Here the volume-averaged effective distance $D_V$ is defined by
\e
\label{dv}
D_V(z) \equiv c\left [(1+z)^2D_A^2(z)\frac {z} {H(z)}\right]^{\frac 1 3}.
\q
The baryon drag epoch $z_d$ is given by Eisenstein \& Hu \cite{Eisenstein:1997ik}, i.e.
\e
\label{zd}
z_d = \frac {1291(\Omega_m h^2)^{0.251}}{1+0.659(\Omega_m h^2)^{0.828}}[1+b_1(\Omega_b h^2)^{b_2}]\ ,
\q
where
\m
b_1 &=& 0.313(\Omega_m h^2)^{-0.419}[1+0.607(\Omega_m h^2)^{0.674}]\ , \\
b_2 &=& 0.238(\Omega_m h^2)^{0.223}\ .
\n

\begin{figure}[H]
\begin{center}
\includegraphics[width=15 cm]{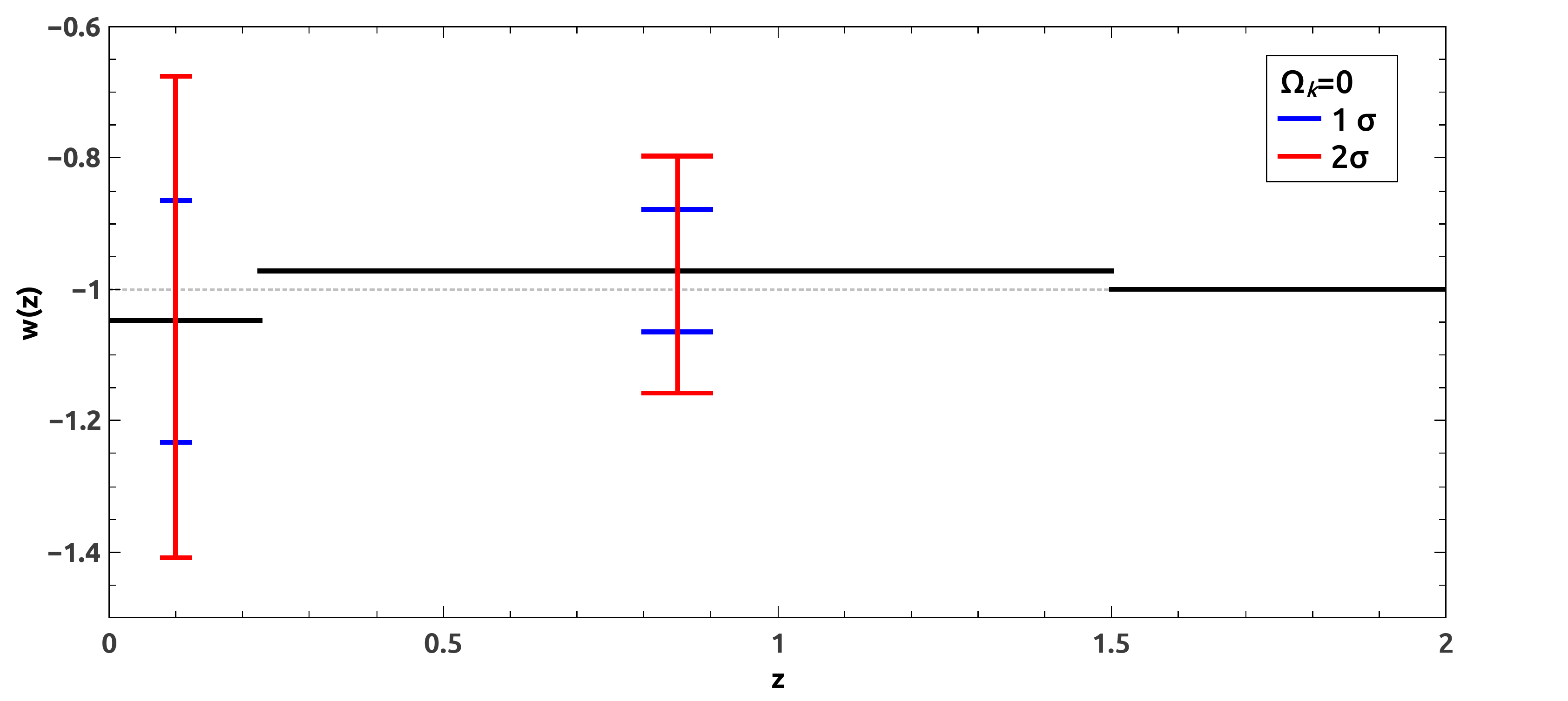}
\end{center}
\caption{EOS of FBDE in spatially flat universe.}
\label{wbin}
\end{figure}

\begin{figure}[H]
\begin{center}
\includegraphics[width=10 cm]{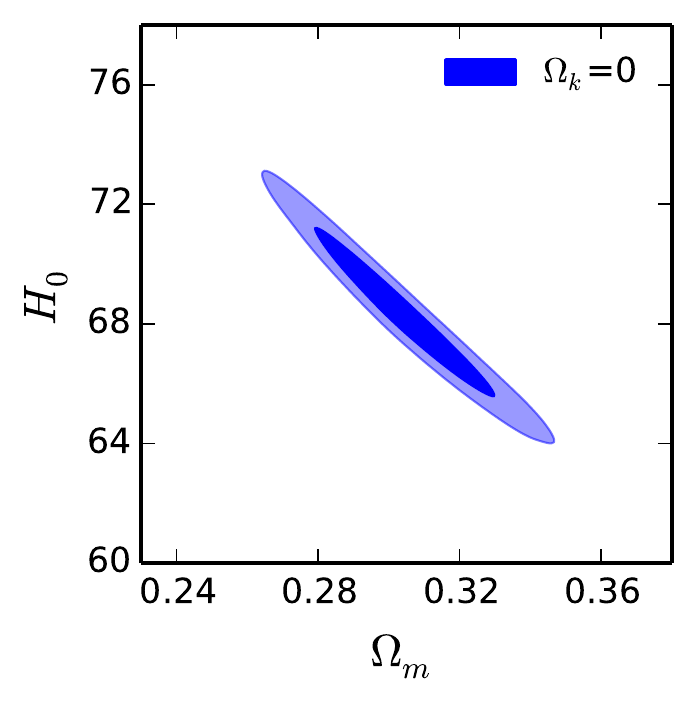}
\end{center}
\caption{Marginalized 2D contours of 1 $\sigma$ and 2 $\sigma$ C.L. in $H_0-\Omega_m$ plane for FBDE given by a combination of distance priors, Union 2.1 and low redshift BAO data.}
\label{test_newdistance_bin_2D}
\end{figure}

In this section, we consider a freely binned dark energy (FBDE) model proposed in \cite{Huang:2009rf,Li:2011wb}. In principle, we can consider a freely N-binned dark energy model. Compared to $\Lambda$CDM model, there are more additional free parameters, i.e. the binning redshift points and EOS of each bin: $z_1, ..., z_i, ..., z_{N-1}$ and $w_1, ..., w_i, ..., w_N$. Unfortunately, it will be very time-consuming if all of these parameters are allowed to vary freely in our full MCMC analysis. Therefore, we will first find a set of $\{z_i\}$ to minimize $\chi^2$ in our analysis, and then we fix $\{z_i\}$ to do the MCMC sampling of the reduced parameter space. In this paper, we deal with a freely 3-binned dark energy model for simplicity. For the first redshift span ($z\leq z_1$), the EOS of DE is denoted by $w_1$; for the second redshift span ($z_1<z\leq z_2$), the EOS of DE is $w_2$. Since at higher redshifts $(z>z_2)$ the energy density of DE become quite small compared to that of matter, the expansion of the universe is insensitive to DE and the EOS of DE can be fixed as $-1$ for convenience. Meanwhile, we take $z_2=1.5$, but $z_1$($>0.1111$) is determined by a best-fit analysis which can yield a minimal $\chi^2$. In this FBDE model, we have
\begin{align}\label{}
\frac {\rho_{de}(z)}{\rho_{de}(0)}=
\begin{cases}
(1+z)^{3(1+w_1)}, ~~z\leq z_1\ ;\\
(1+z_1)^{3(1+w_1)}\left(\frac{1+z}{1+z_1}\right)^{3(1+w_2)}, ~~z_1<z\leq z_2\ ;\\
(1+z_1)^{3(1+w_1)}\left(\frac{1+z_2}{1+z_1}\right)^{3(1+w_2)}, ~~z>z_2\ .
\end{cases}
\end{align}
Now we can adopt the combined datasets of distance priors, Union~2.1 and low redshift BAO to constrain FBDE model with free parameters $\{\Omega_m, H_0, w_1, w_2, \Omega_b h^2, z_1\}$. We first set the option ``action=2" in \textsf{CosmoMC} to minimize $\chi^2$ and we get $z_1=0.2256$. Then we fix $z_1=0.2256$ and set the option ``action=0'' to do a MCMC sampling and we constrain the other cosmological parameters. The EOS of FBDE is showed in Fig.~\ref{wbin} where $w_1=-1.0473\pm0.1852$ and $w_2=-0.9720\pm0.0932$ at $68\%$ C.L. which are consistent with $w=-1$ within $1\sigma$ C.L.. Moreover, the  constraints on $\Omega_m$ and $H_0$ are showed in Fig.~\ref{test_newdistance_bin_2D}. Maginallizing over other parameters, we get $H_0=68.34\pm1.83~\rm{km/s\cdot Mpc^{-1}}$ and $\Omega_m=0.3043\pm0.0165$ at $1\sigma$ C.L..

\begin{figure}[H]
\begin{center}
\includegraphics[width=15 cm]{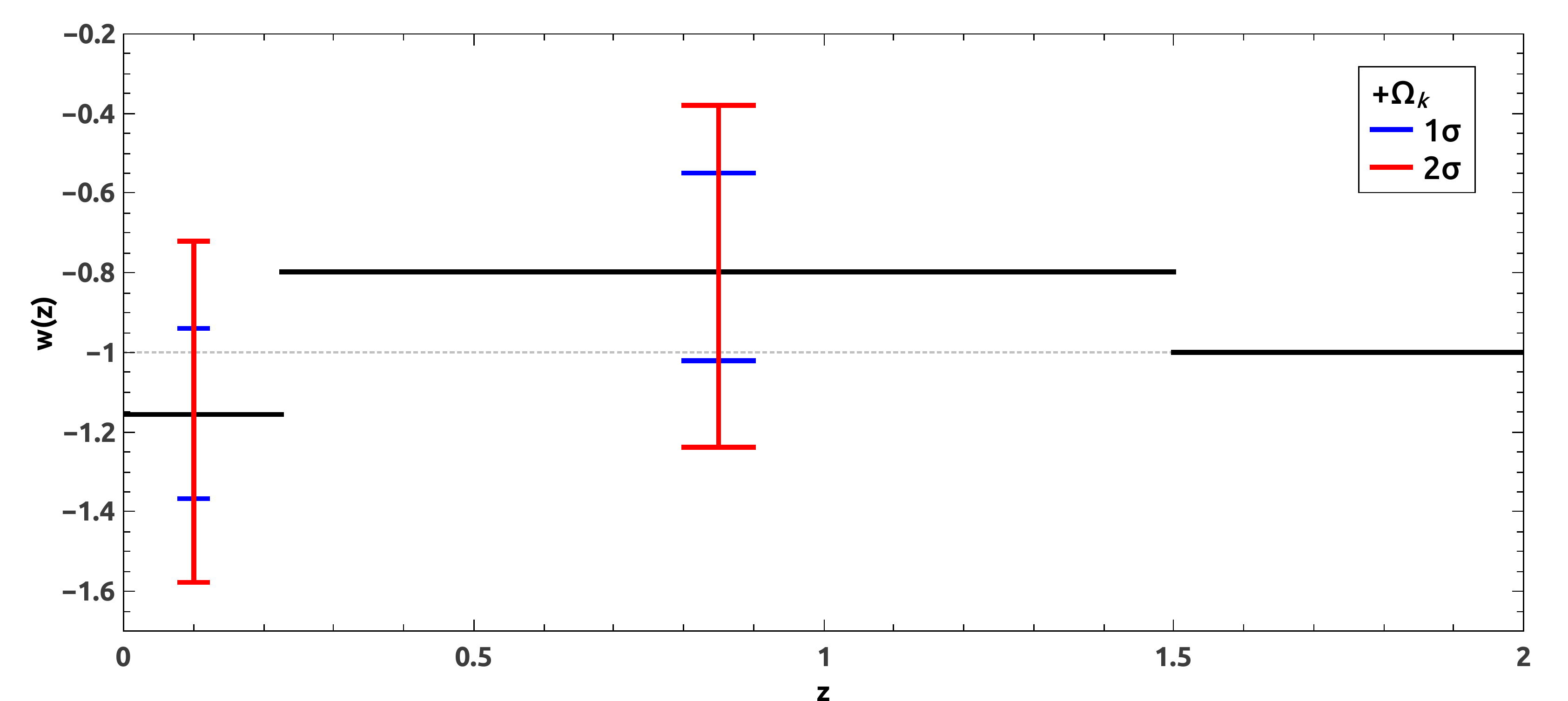}
\end{center}
\caption{EOS of FBDE with spatial curvature.}
\label{wbink}
\end{figure}

\begin{figure}[H]
\begin{center}
\includegraphics[width=10 cm]{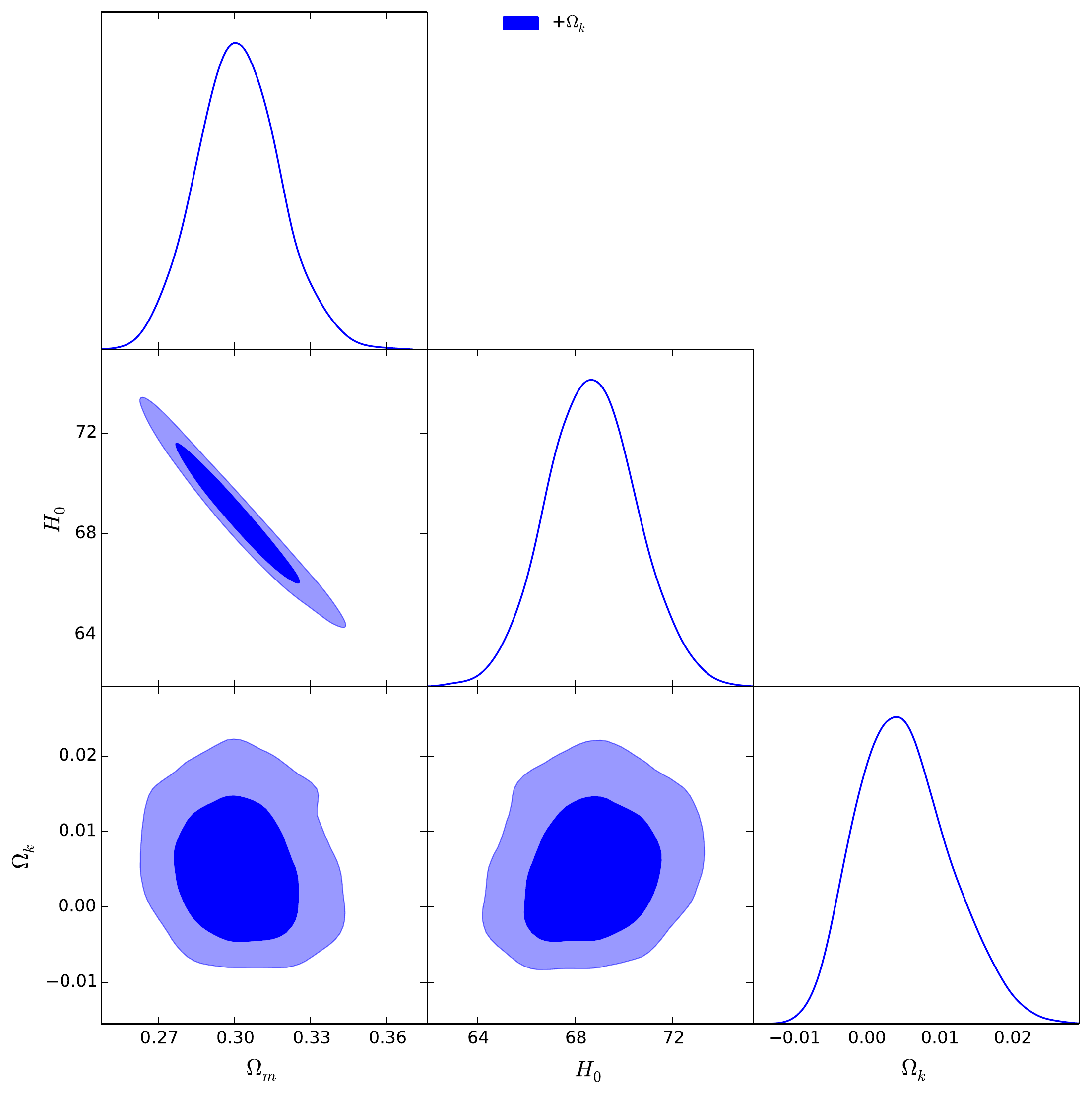}
\end{center}
\caption{Marginalized two-dimension probability contours and one-dimension probability distribution functions of $\Omega_m$, $H_0$ and $\Omega_k$ for FBDE+$\Omega_k$ model given by a combination of distance priors, Union 2.1 and BAO measurements at low redshift.}
\label{test_newdistance_bink_tri}
\end{figure}

Here we are also interested in how the former results change in the FBDE model once the spatial curvature is taken into account, namely the FBDE+$\Omega_k$ model. Similar to the case with $\Omega_k=0$, we get $z_1=0.2257$ and then fix it to figure out the constraints on the other parameters $\{\Omega_m, H_0, \Omega_k, w_1, w_2, \Omega_b h^2\}$. Our results show up in Figs.~\ref{wbink} and \ref{test_newdistance_bink_tri}. Here $w_1=-1.1560\pm0.2178$ and $w_2=-0.7977\pm0.2234$ at $1\sigma$ C.L., which are also consistent with the prediction of $\Lambda$CDM model. The constraints on the other parameters are $\Omega_m=0.3017\pm0.0160$, $H_0=68.69\pm1.80~\rm{km/s \cdot Mpc^{-1}}$ and $\Omega_k=0.0053\pm0.0062$ at $1\sigma$ C.L.. A spatially flat universe is included within $68\%$ C.L..

\subsection{BAO at $z=2.34$}

The BAO feature at high redshift can be measured through absorption in Ly$\alpha$ forest (Ly$\alpha$F) \cite{bao3}.
According to \cite{Delubac:2014aqe,Font-Ribera:2013wce}, $1/H(z=2.34)$ predicted by the $\Lambda$CDM model constrained by \emph{Planck} 2013 data is approximately 2$\sigma$ lower than that measured by the Ly$\alpha$F auto-correlation, while $D_A(z=2.36)$ is approximately 2$\sigma$ larger than that from the Ly$\alpha$F-QSO cross correlation. Furthermore, $c/(H(z=2.34)r_s(z_d))=9.18\pm0.28$ and $9.15^{+0.20}_{-0.21}$ from Ly$\alpha$F auto-correlation and Combined Ly$\alpha$F, respectively. However, the $\Lambda$CDM constrained by \emph{Planck} 2015 data \cite{Ade:2015xua} predicts $c/(H(z=2.34)r_s(z_d))=8.586\pm0.021$ which is more than $2\sigma$ C.L. lower than those from both Ly$\alpha$F auto-correlation and Combined Ly$\alpha$F.

\begin{figure}[H]
\begin{center}
\includegraphics[width=10 cm]{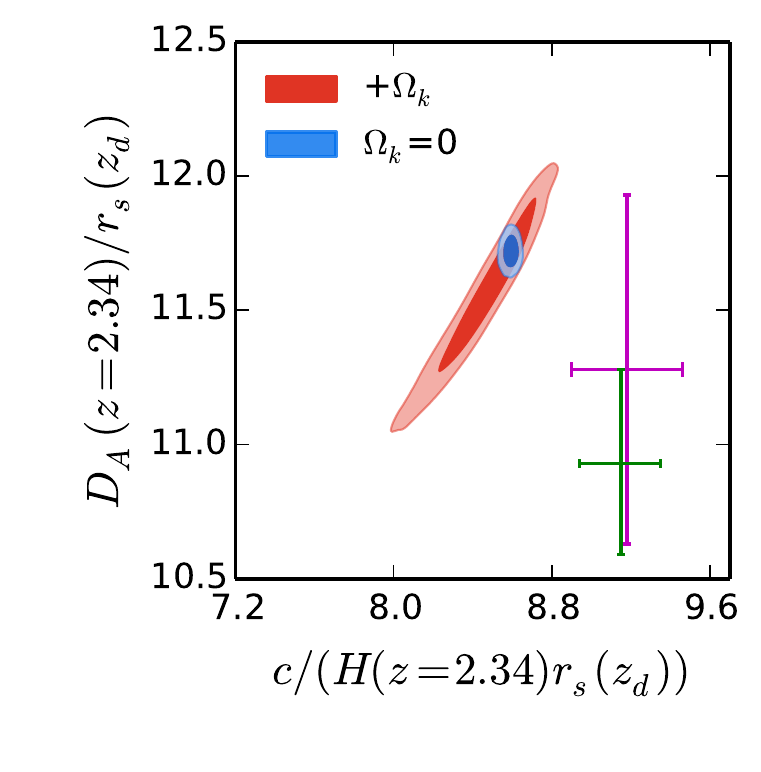}
\end{center}
\caption{The 68\% and 95\% constraints on $D_A(z=2.34)/r_s(z_d)$ and $c/(H(z=2.34)r_s(z_d))$ from FBDE (red) and FBDE+$\Omega_k$ (blue). The purple and green crossed denote the measurements from Ly$\alpha$ auto-correlation and Combined Ly$\alpha$ at $1\sigma$ C.L. respectively.}
\label{answer_baok_2D}
\end{figure}

\begin{figure}[H]
\begin{center}
\includegraphics[width=6.6 cm]{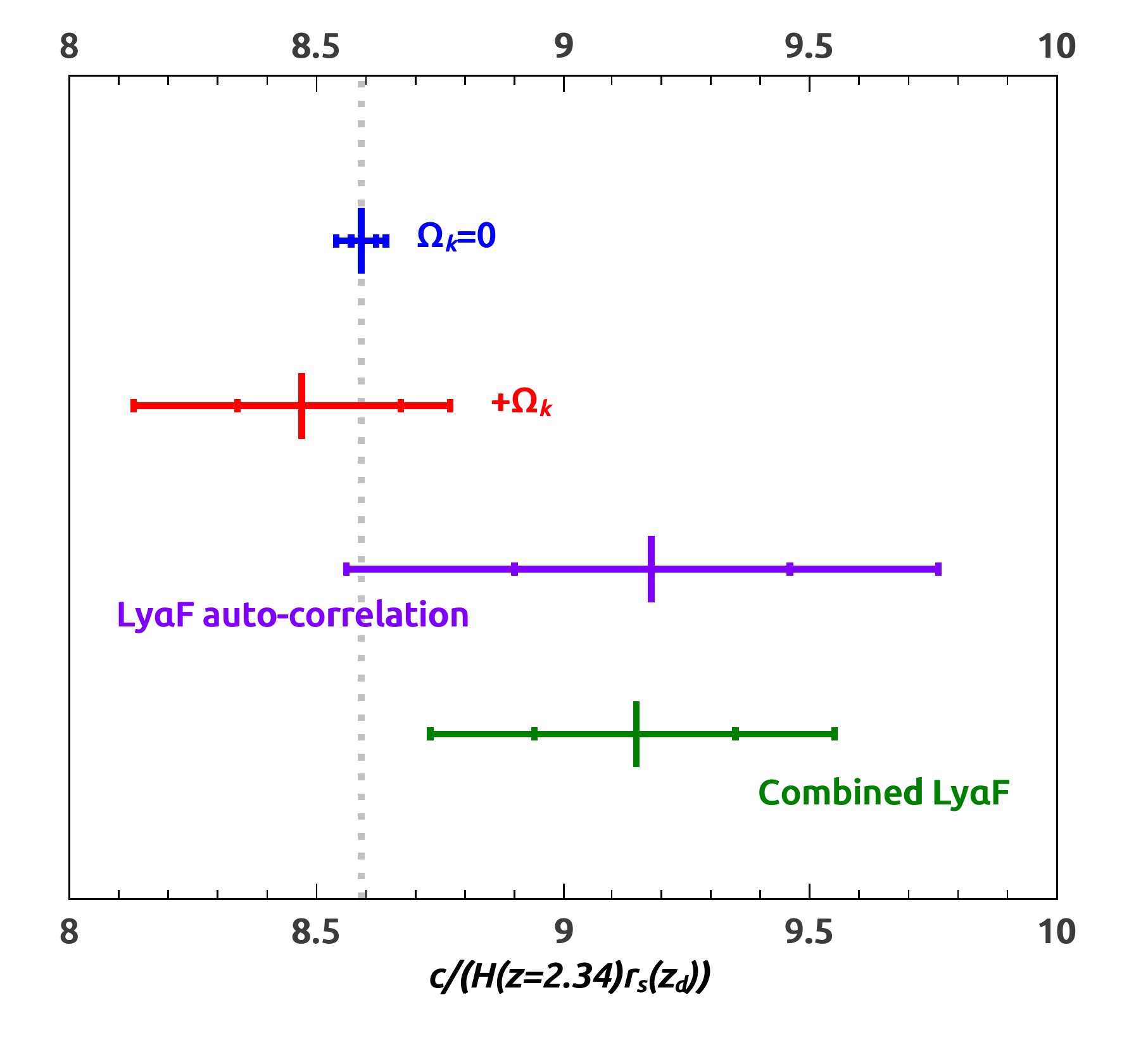}\quad
\includegraphics[width=6.8 cm]{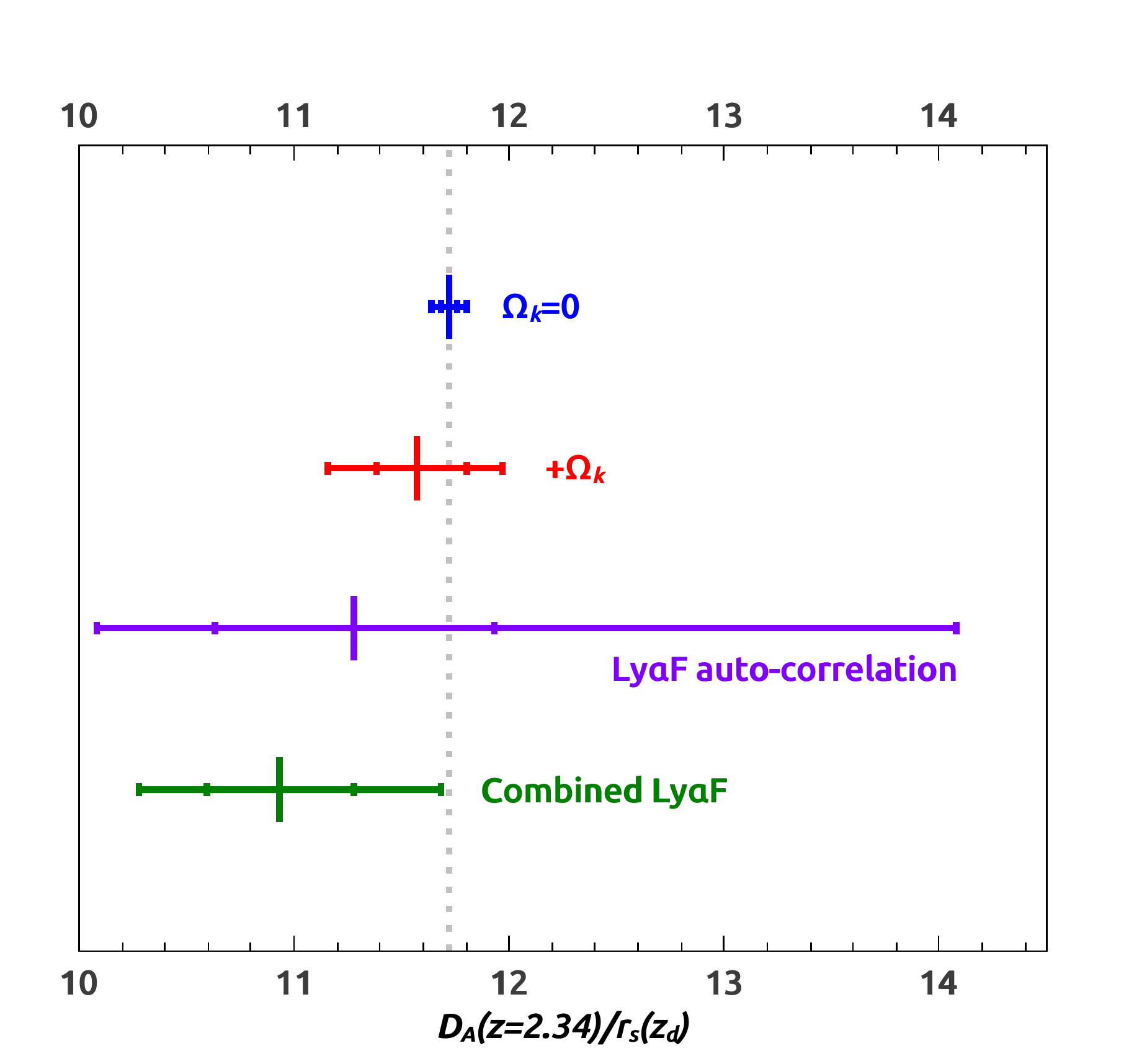}
\end{center}
\caption{Comparison between measurements and our model predictions of $c/(H(z=2.34)r_s(z_d))$ and $D_A(z=2.34)/r_s(z_d)$.}
\label{dadh}
\end{figure}

Here we wonder whether the tension between \emph{Planck} 2015 data and the BAO at $z=2.34$ from Ly$\alpha$F is relaxed in the FBDE and FBDE+$\Omega_k$ models, respectively. The two dimensional contours on $D_A(z=2.34)/r_s(z_d)$ and $c/(H(z=2.34)r_s(z_d))$ predicted by the FBDE and FBDE+$\Omega_k$ models are given in Fig.~\ref{answer_baok_2D}, where the purple and green crosses correspond to Ly$\alpha$F auto-correlation and Combined Ly$\alpha$F, respectively. The marginalized $D_A(z=2.34)/r_s(z_d)$ and $c/(H(z=2.34)r_s(z_d))$ are illustrated in Fig.~\ref{dadh}. The measurements of Ly$\alpha$F auto-correlation and Combined Ly$\alpha$F and the predictions of FBDE and FBDE+$\Omega_k$ models are summarized in Tab.~\ref{Lya}.

To summarize, FBDE and FBDE$+\Omega_k$ models give consistent predictions on both $D_A(z=2.34)/r_s(z_d)$ and $c/(H(z=2.34)r_s(z_d))$. But $c/(H(z=2.34)r_s(z_d))$ predicted by FBDE and FBDE$+\Omega_k$ models are still lower than that from Ly$\alpha$F auto-correlation measurement at around $2\sigma$ C.L., and become even worse compared to that from Combined Ly$\alpha$F. Even though $D_A(z=2.34)/r_s(z_d)$ predicted by FBDE and FBDE$+\Omega_k$ models are consistent with that from Ly$\alpha$ auto-correlation measurement, there is still an around 2$\sigma$ discrepancy compared to Combined Ly$\alpha$F measurement. In a word, relaxing the dark energy model from the cosmological constant and adding the spatial curvature cannot significantly relax the tension between \emph{Planck} 2015 data and the BAO at $z=2.34$ from Ly$\alpha$F.

\begin{table*}[!htp]
\centering
\renewcommand{\arraystretch}{1.5}
\begin{tabular}{ccccc}
\hline\hline
Measurement & Parameter            & mean    &1$\sigma$           & 2$\sigma$ \\
\hline
Ly$\alpha$ auto-correlation  &$D_A(z=2.34)/r_s(z_d)$  &$11.28$  &$^{+0.65}_{-0.65}$  &$^{+2.80}_{-1.20}$ \\

                             &$c/(H(z=2.34)r_s(z_d))$ &$9.18$   &$^{+0.28}_{-0.28}$  &$^{+0.60}_{-0.60}$ \\
\hline
Combined Ly$\alpha$          &$D_A(z=2.34)/r_s(z_d)$  &$10.93$  &$^{+0.35}_{-0.34}$  &$^{+0.75}_{-0.65}$ \\

                             &$c/(H(z=2.34)r_s(z_d))$ &$9.15$   &$^{+0.20}_{-0.21}$  &$^{+0.40}_{-0.42}$ \\
\hline
\hline
Model & Parameter & mean    &1$\sigma$           & 2$\sigma$ \\
\hline
FBDE                         &$D_A(z=2.34)/r_s(z_d)$  &$11.72$  &$^{+0.04}_{-0.04}$  &$^{+0.08}_{-0.08}$ \\

                             &$c/(H(z=2.34)r_s(z_d))$ &$8.59$   &$^{+0.03}_{-0.02}$  &$^{+0.05}_{-0.05}$ \\
\hline
FBDE+$\Omega_k$              &$D_A(z=2.34)/r_s(z_d)$  &$11.57$  &$^{+0.23}_{-0.19}$  &$^{+0.40}_{-0.41}$ \\

                             &$c/(H(z=2.34)r_s(z_d))$ &$8.47$   &$^{+0.20}_{-0.13}$  &$^{+0.30}_{-0.34}$ \\
\hline
\end{tabular}
\caption{The measurements and predictions of $c/(H(z=2.34)r_s(z_d))$ and $D_A(z=2.34)/r_s(z_d)$.}
\label{Lya}
\end{table*}

\section{Summary and discussion}

We update the distance priors by utilizing the CMB data, especially including high $\ell$ polarizations data, recently released by the \emph{Planck} Collaboration. Compared to those given in \cite{Ade:2015rim} where only \emph{Planck} TT+LowP data were used, our results impose at least $30\%$ tighter constraints. Combining the distance priors given in this paper with Union~2.1 and low redshift BAO datasets, we constrain the cosmological parameters in both FBDE and FBDE$+\Omega_k$ models and find that the $\Lambda$CDM model in a spatially flat universe is included within $1\sigma$ C.L..

In the literatures, there are debates on the tensions on $D_A(z=2.34)/r_s(z_d)$ and $c/(H(z=2.34)r_s(z_d))$ between  \emph{Planck} 2015 data and the BAO at $z=2.34$ from Ly$\alpha$F in the $\Lambda$CDM model. We find that these tensions cannot be relaxed in both FBDE and FBDE$+\Omega_k$ models. Of course, it is also possible that there are some unknown systemic errors in the data analysis for CMB and/or Ly$\alpha$F. It is always worthy investigating all of these possibilities in the near future.

\vspace{5mm}
\noindent{\large \bf Acknowledgments}

We would like to thank C.~Cheng for useful conversation. This work is supported by Top-Notch Young Talents Program of China and grants from NSFC (grant NO. 11322545, 11335012 and 11575271).


\clearpage

\end{document}